# AR Glulam: Accurate Augmented Reality Using Multiple Fiducial Markers for Glulam Fabrication

**Alexander Htet Kyaw**
MIT

**Arvin Xu**
Cornell University

**Sasa Zivkovic**
Cornell University

**Gwyllim Jahn**
Fologram

**Cameron Newnham**
Fologram

**Nick Van Den Berg**
Fologram

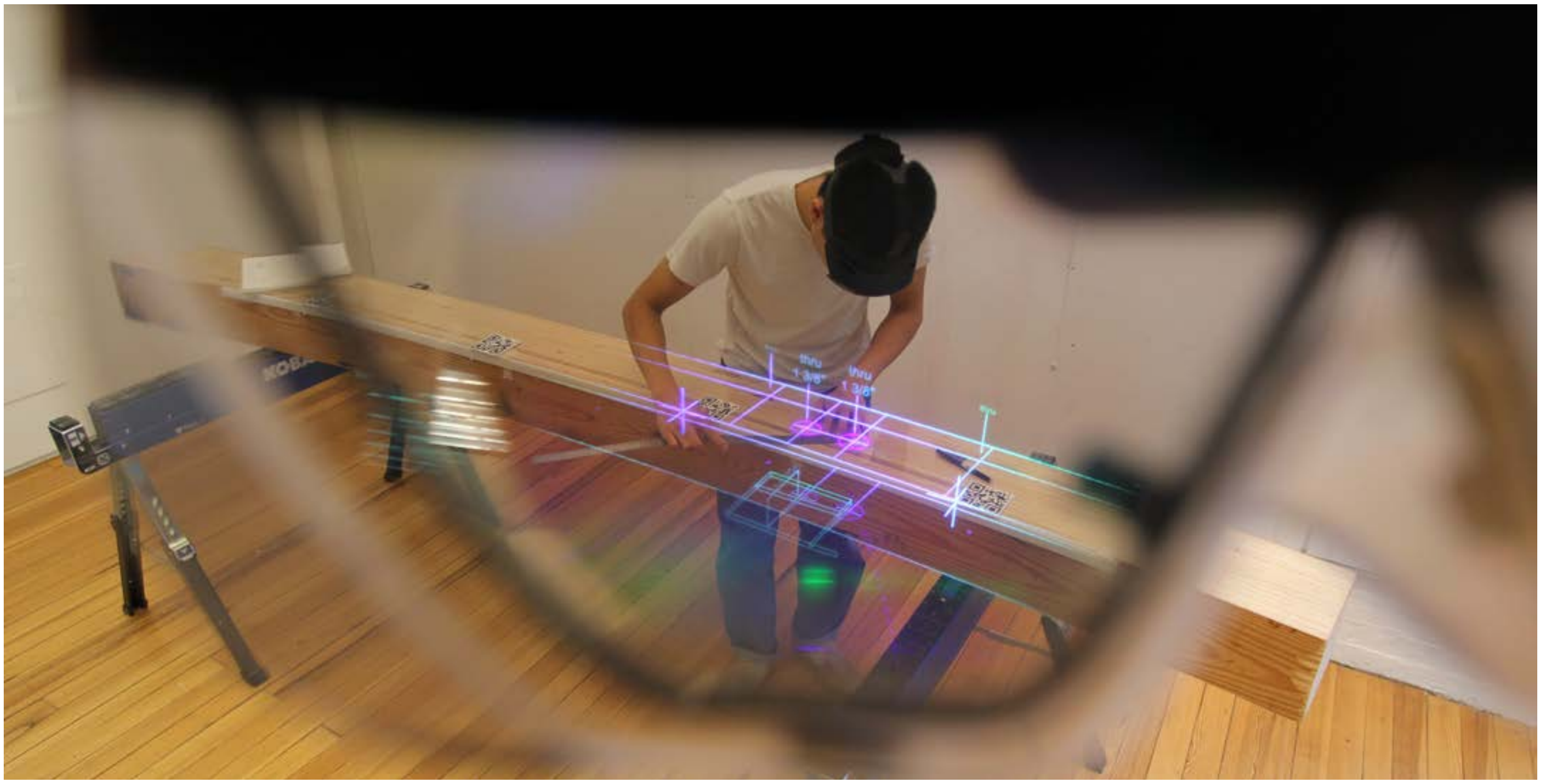

**1** Multiple Fiducial Marker for High Precision AR in a Lab Environment

## Introduction

Recent advancements in Augmented Reality (AR) have demonstrated applications in architecture, design, and fabrication (Song, Koeck, and Luo 2021). Compared to conventional 2D construction drawings, AR can be used to superimpose contextual instructions, display 3D spatial information and enable on-site engagement. Despite the potential of AR, the widespread adoption of the technology in the industry is limited by its precision (Chi, Kang, and Wang 2013).

Precision is important for projects requiring strict construction tolerances, design fidelity and fabrication feedback. For example, the manufacturing of glulam beams requires tolerances of less than 2mm (Jones and Brischke 2017). The goal of this project is to explore the industrial application of using multiple fiducial markers for high-precision AR fabrication (Figure 1). While the method has been validated in lab settings with a precision of 0.97 (Figure 2), this paper focuses on fabricating glulam beams in a factory setting with an industry manufacturer, *Unalam Factory* (Kyaw, et al, 2023).

**PRODUCTION NOTES**

Client: Unalam Factory

Status: Completed

Location: Pennsylvania, New Hampshire, Connecticut, USA

Date: 2022

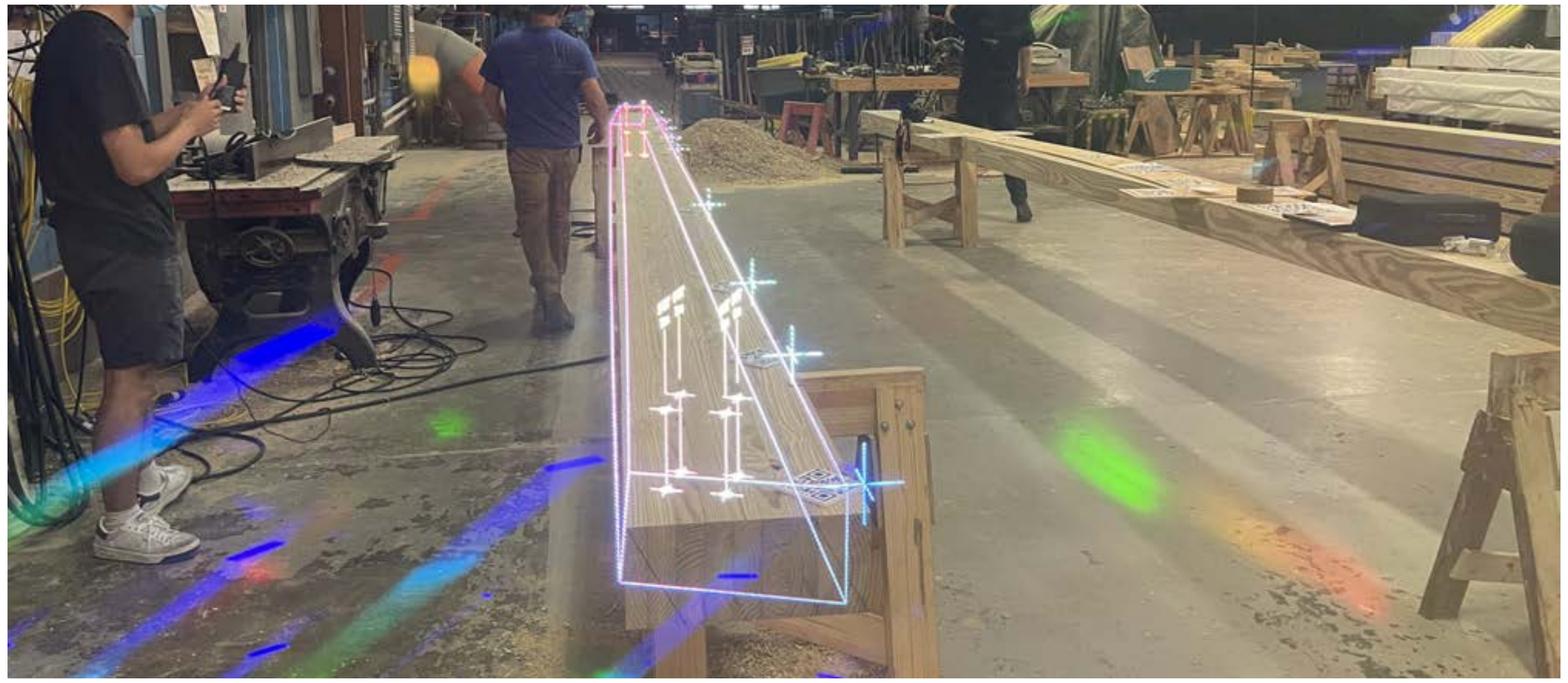

2 Multiple Fiducial Marker for High Precision AR in a Factory Environment

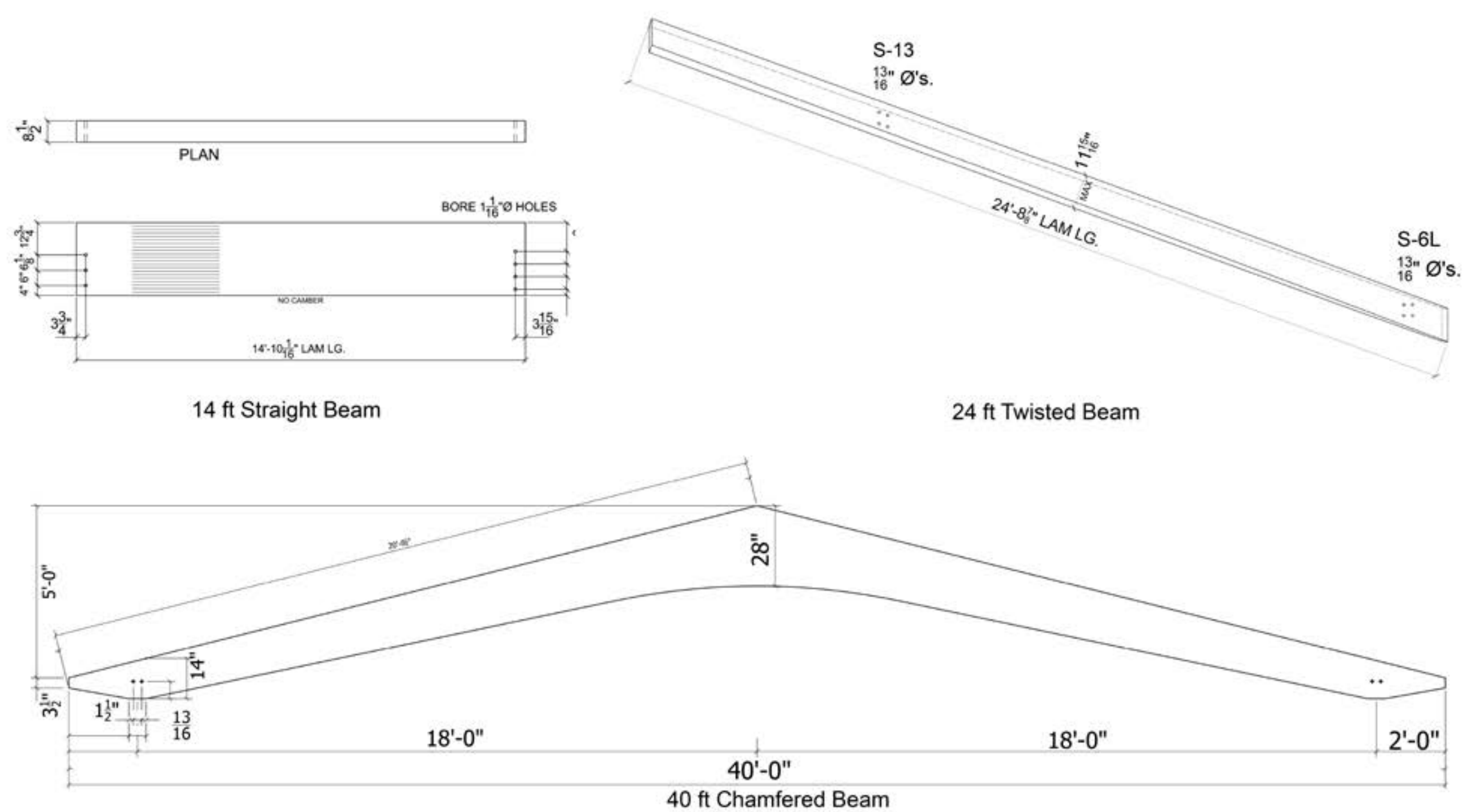

3 Three Different Types of Beams with Varying Shapes, Dimensions and

**Augmented Reality Software and Equipment**

This project utilizes the *Twinbuild* software to recognize multiple quick response (QR) fiducial markers for correcting drift errors in AR. Each QR marker is linked to know points in the physical environment or on the beam. These points are interpolated and serve as reference points to maintain alignment accuracy using the HoloLens2 AR headset.

**Experiment Setup**

Three types of beams were fabricated: a 14-foot straight beam for the Upper Merion Area High School in Pennsylvania, a 24-foot twisted beam  for the Milford band shed in New Hampshire, and a 40-foot cable roof beam for the Winding Trails Pavilion in Connecticut (Figure 3). The 14-foot beam have five markers at 2.5-foot intervals along the beam edge (Figure 4). The 24-foot twisted beam have

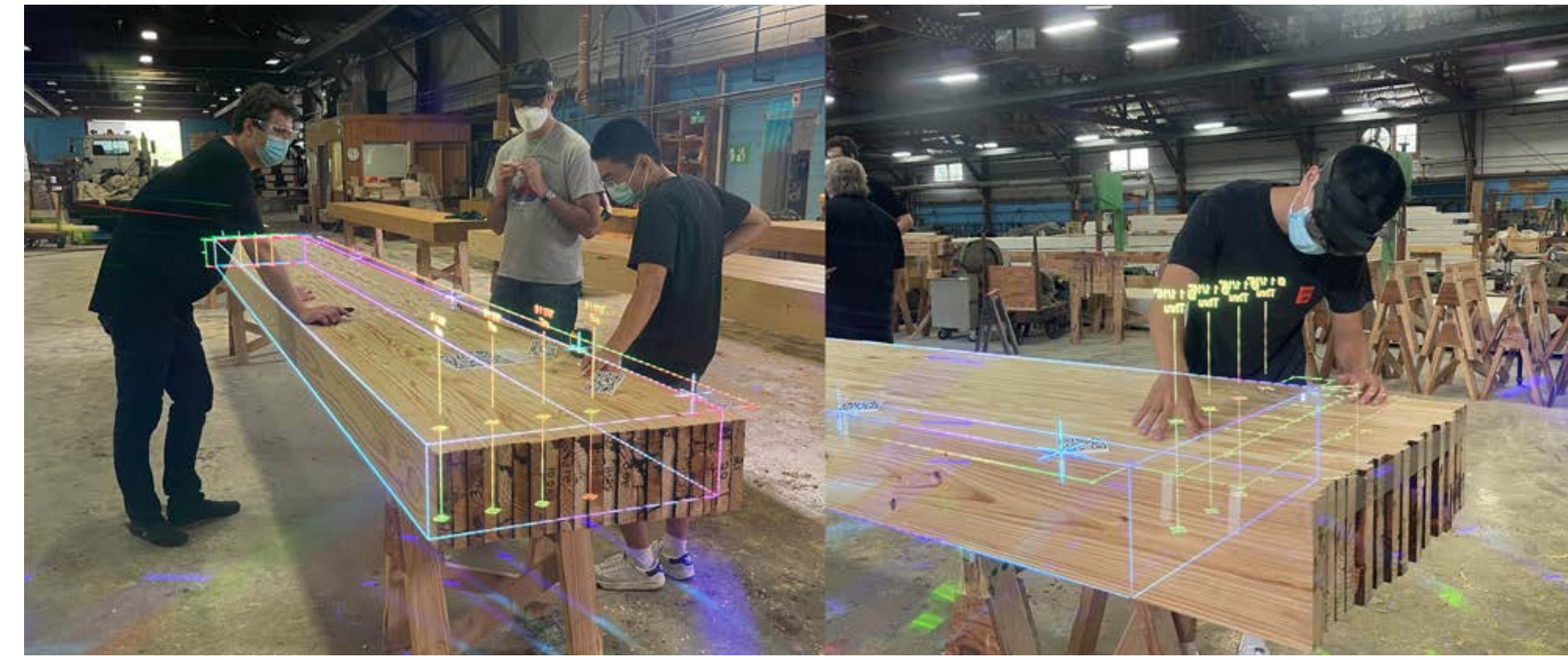

4 Fourteen Foot Long Twisted Beam for the Upper Merion Area High School in Pennsylvania

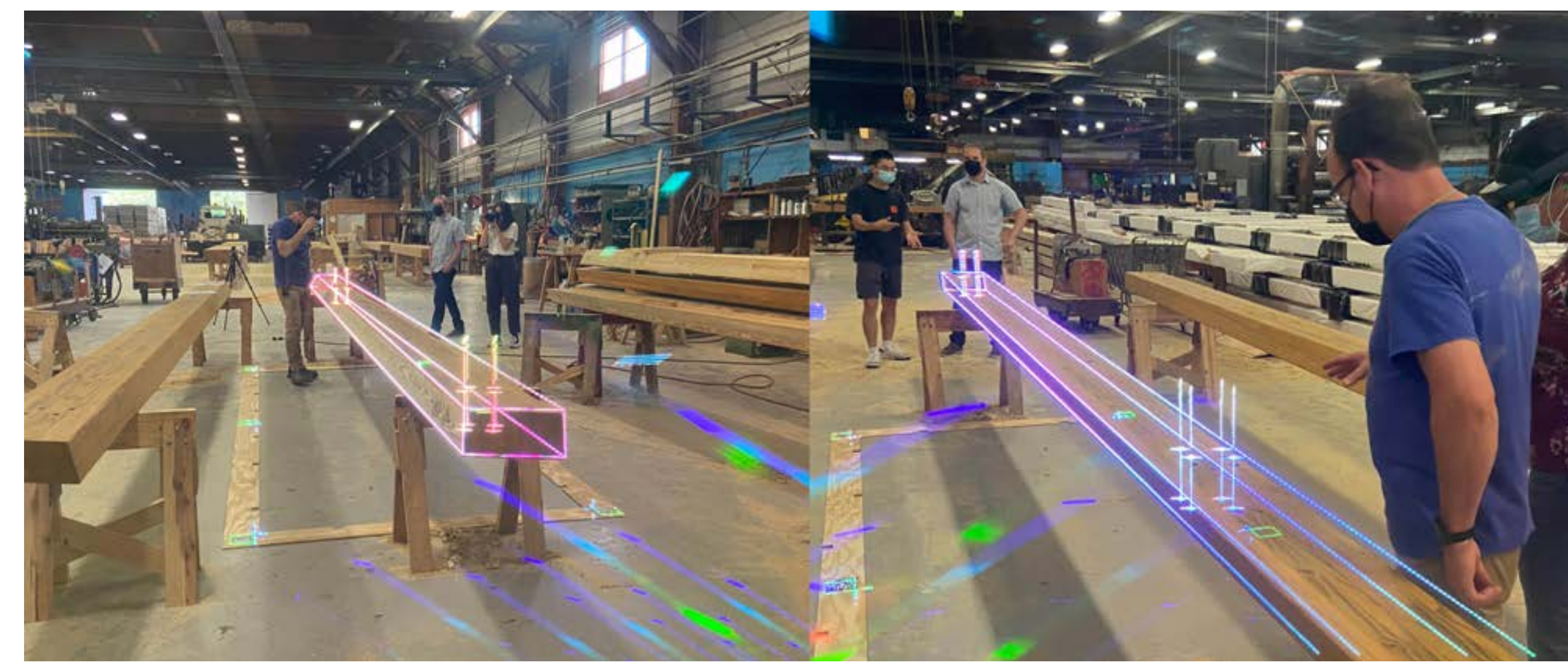

5 Twenty-Four Foot Long Twisted Beam for the Milford band shed in New Hampshire

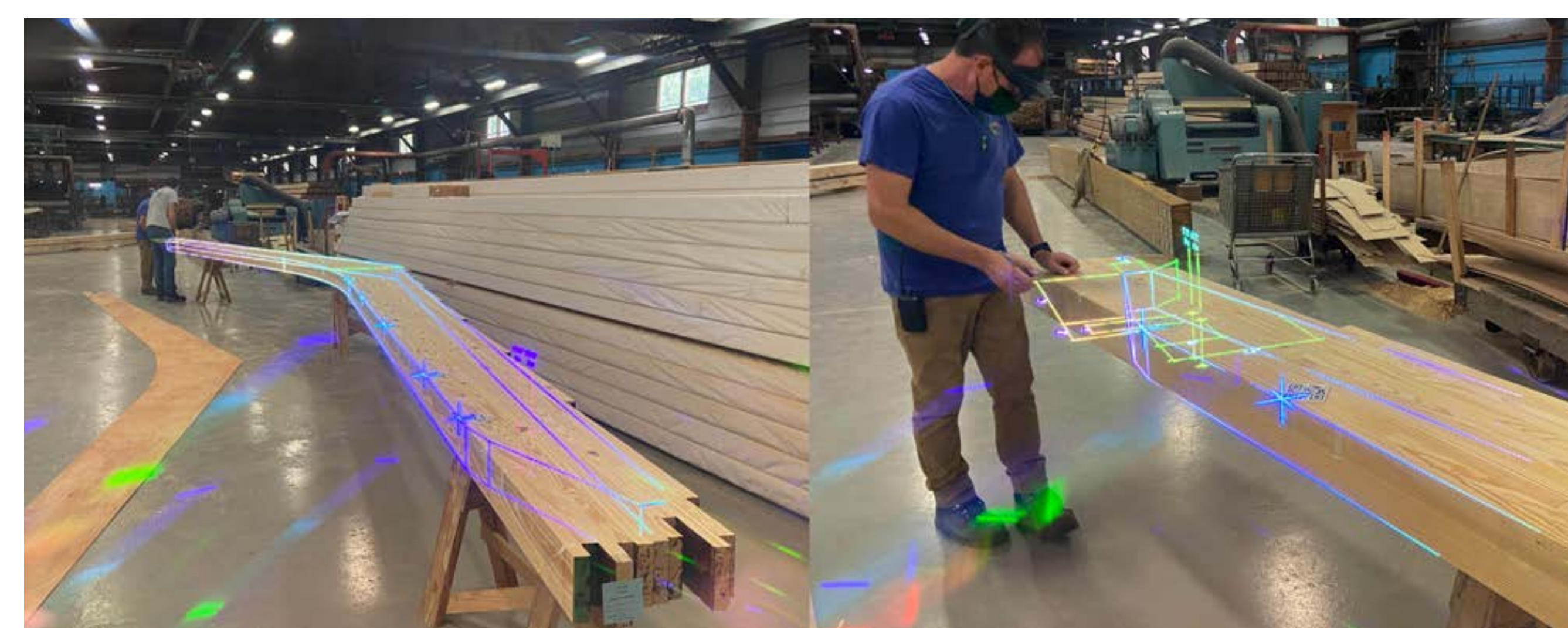

6 . Forty Foot Long Chamfered Beam for the Winding Trails Pavilion in Connecticut

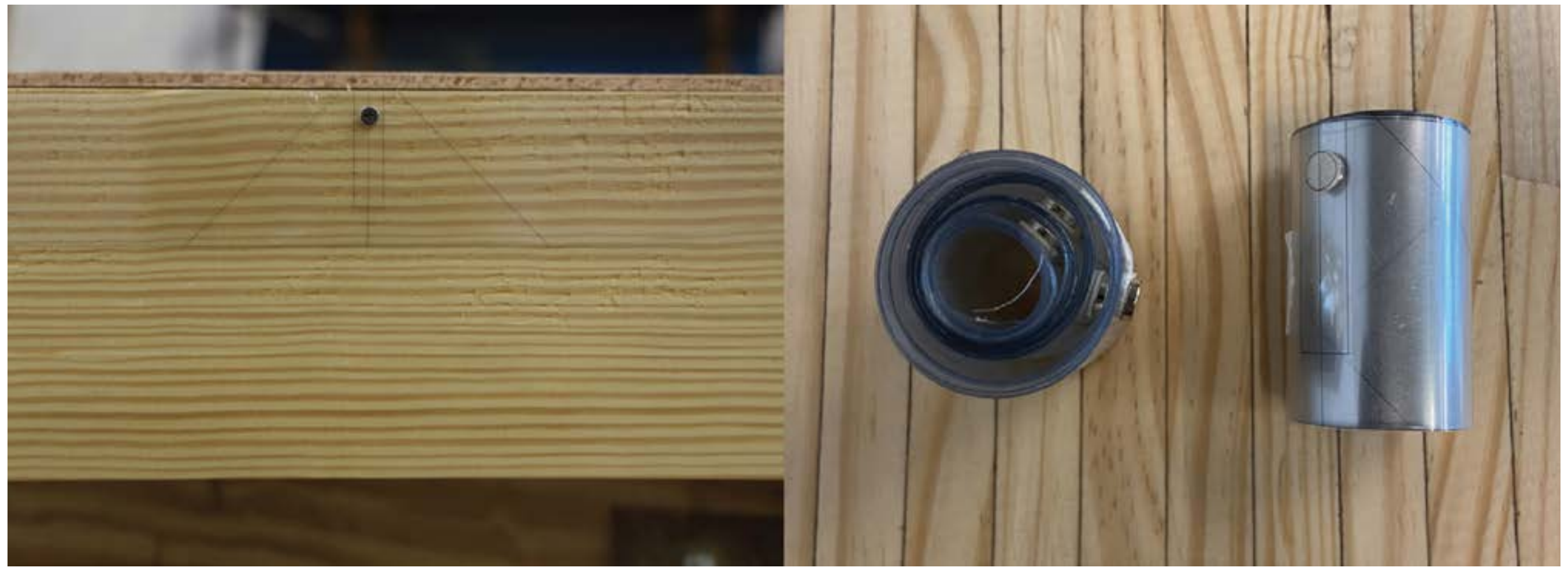

7 Clear Strip with Magnets Used as Guide for Markers

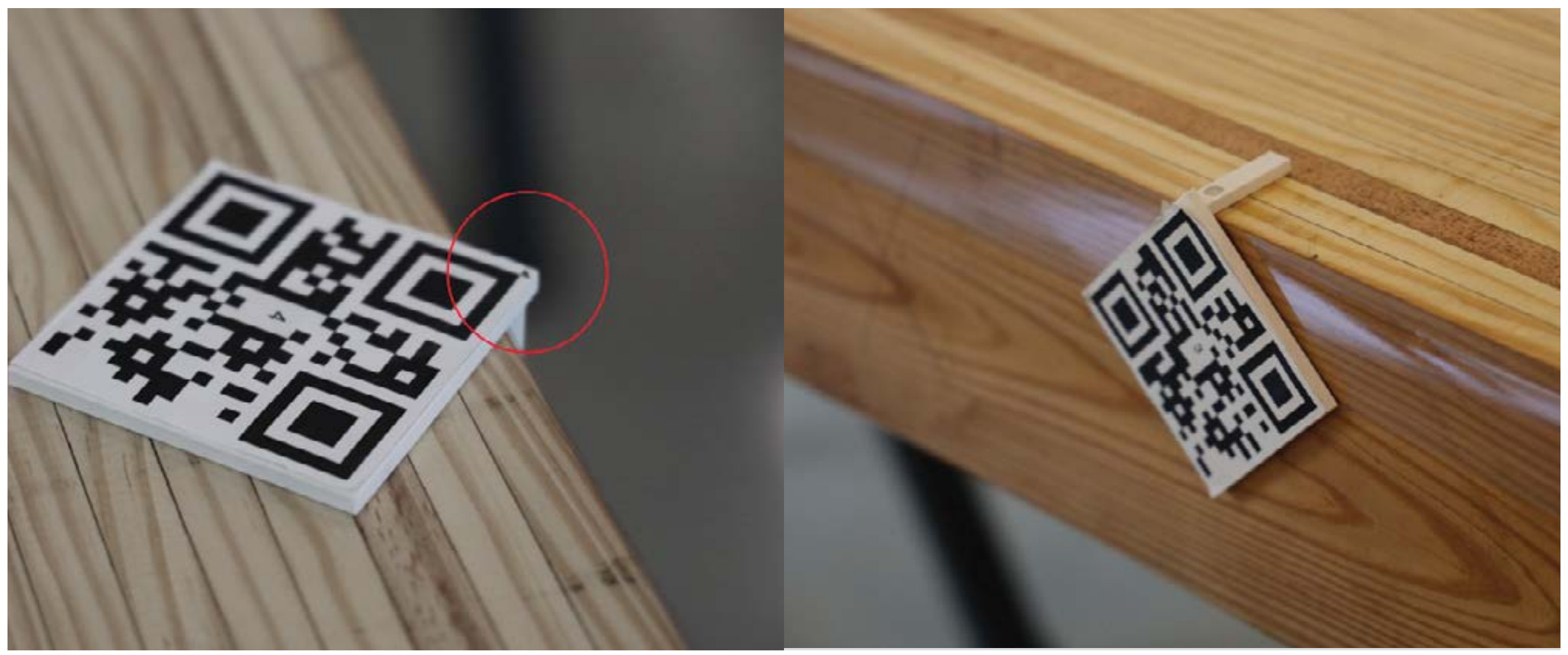

8 Vertical and Horizontal Placement of Markers

6 markers around the beam at 6 feet intervals (Figure 5). Since the edge of the twisted beam is not flat, the markers are placed around the beam. The 40-foot beam have 10 markers at 4-foot interval along the beam edge to reduce the number of markers (Figure 6). The markers arrangements are varied depending on the usage and the type of beam.

### User Guide and Web Interface

A user guide is developed for instructing new users. The procedure involves using a clear strip with magnets spaced at 2.5-foot intervals and 4-foot intervals for easy attachment of markers (Figure 7). This strip is attached to the edge of the beam. Markers are printed on a 3D-printed magnetic holder that can be snapped onto this strip. The markers can be positioned either horizontally on the top-facing edge of the beam or vertically on the side-facing edge (Figure 8).

Each beam is assigned a unique QR code that references its digital model. These models are stored on a server, allowing users to upload them via a web interface (Figure 9a). This eliminates the need for the headset to be connected to a computer for streaming AR instructions. By scanning the QR code, the headset can access and display the corresponding beam model directly from the server. Additionally, the makers can also be printed from the web interface, enabling easy access (Figure 9b).

### Precision in the Factory

Factory measurements indicate that the 14-foot straight beam, with markers positioned every 2.5-foot, has an average deviation of 1.2 mm. The 40-foot beam with markers positioned at every 4-foot, has an average deviation of 1.7 mm. Both are within the 2 mm tolerance standard required for glulam beam fabrication. The 24-foot beam, with markers positioned around the beam every 6 feet, has an average deviation of 2.3 mm. Placing markers at the further apart yield lower accuracy. However, it provides a simpler setup that can be still be effectively used for other processes such as quality control.

Compared to results in the laboratory, the accuracy in a factory setting is slightly lower. This discrepancy is likely attributable to variations in lighting conditions and the presence of moving objects within the factory (Figure 10). Further studies in an industrial setting are necessary to investigate these factors using the multiple marker method. Lastly, while markers can be positioned both horizontally and vertically, users prefer to place markers in the same orientation as the fabrication instructions since it is easier to scan.

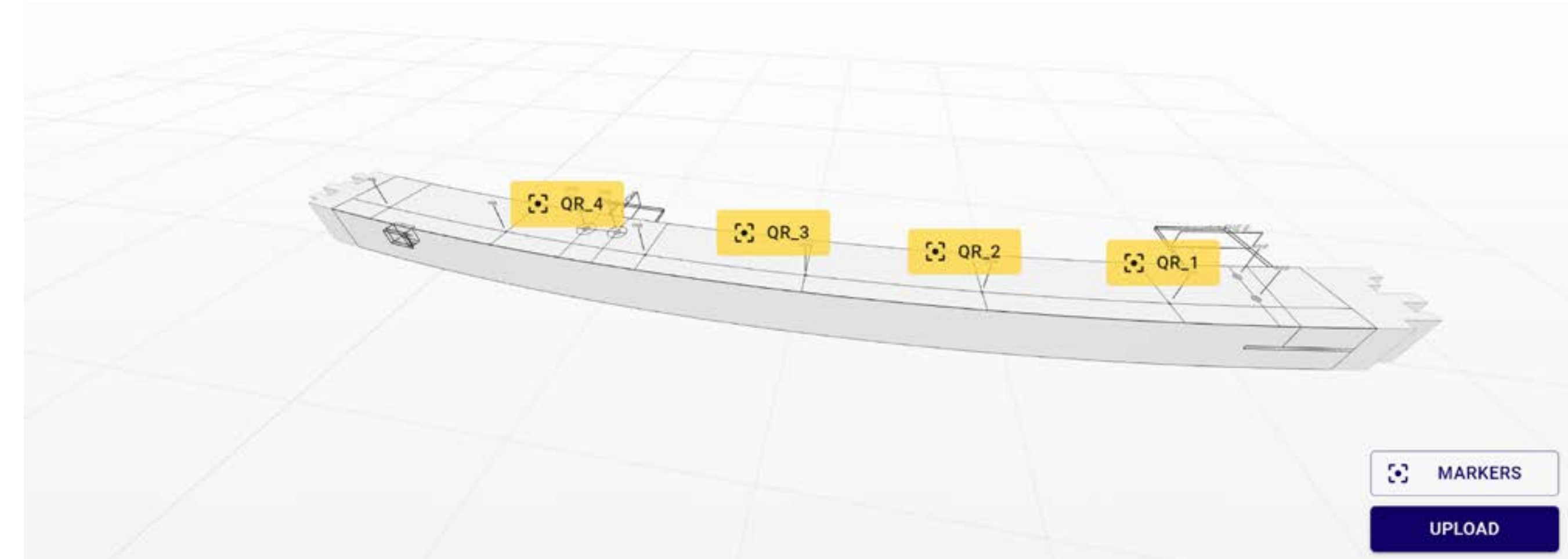

**9a** Web Interface for Uploading Fabrication Model.

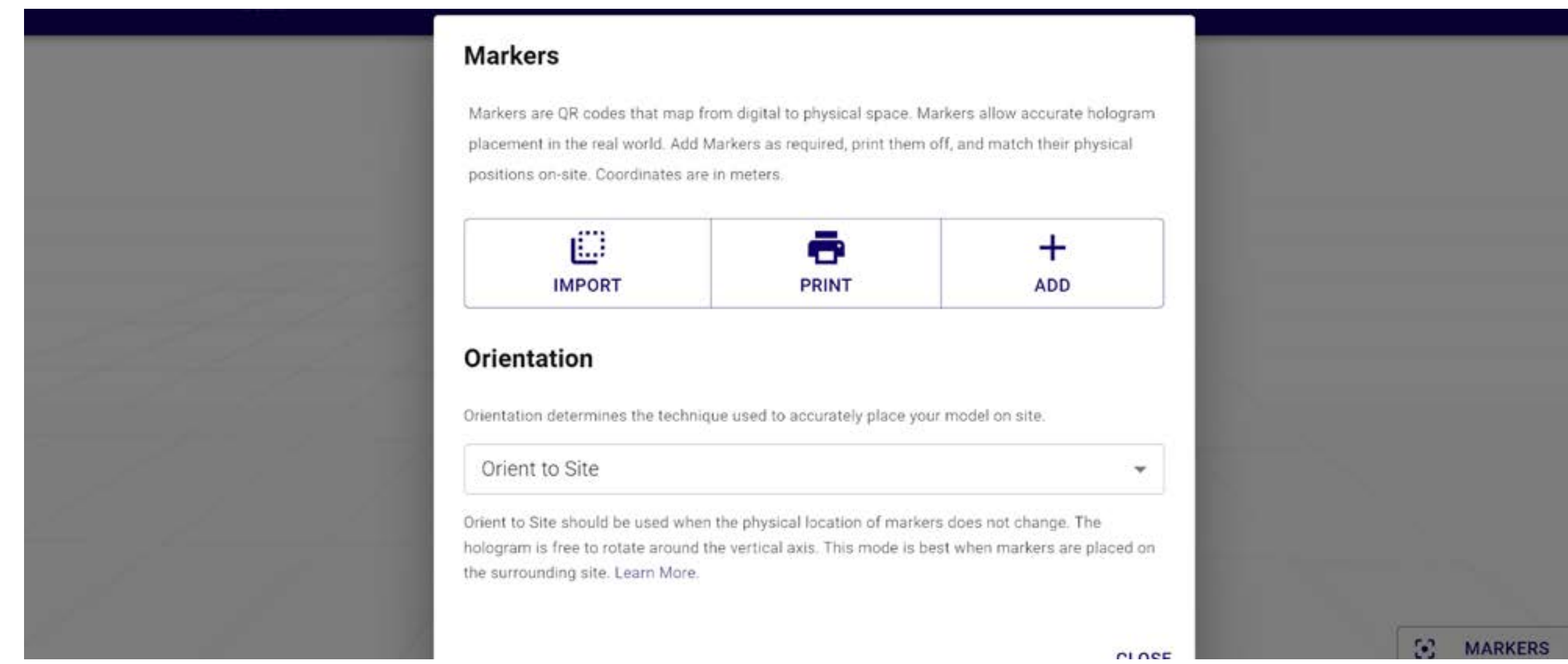

**9a** Web Interface for Printing QR Markers.

### Conclusion

The paper demonstrates the industrial application of multiple fiducial markers for high precision-augmented reality fabrication of glulam beams in a scalable manner (Figure 10). Future studies can use this work to extend its application to larger construction projects. This project challenges the idea that AR is only suitable for low-precision projects, by demonstrating its potential for high-precision applications. With enhanced precision in AR, there are new opportunities for applications where accuracy is important such as feedback-based fabrication (Kyaw at el.2024), human-robot interaction (Kyaw at el.2024), and on-site design fabrication (Kyaw, Otto and Lok 2023).

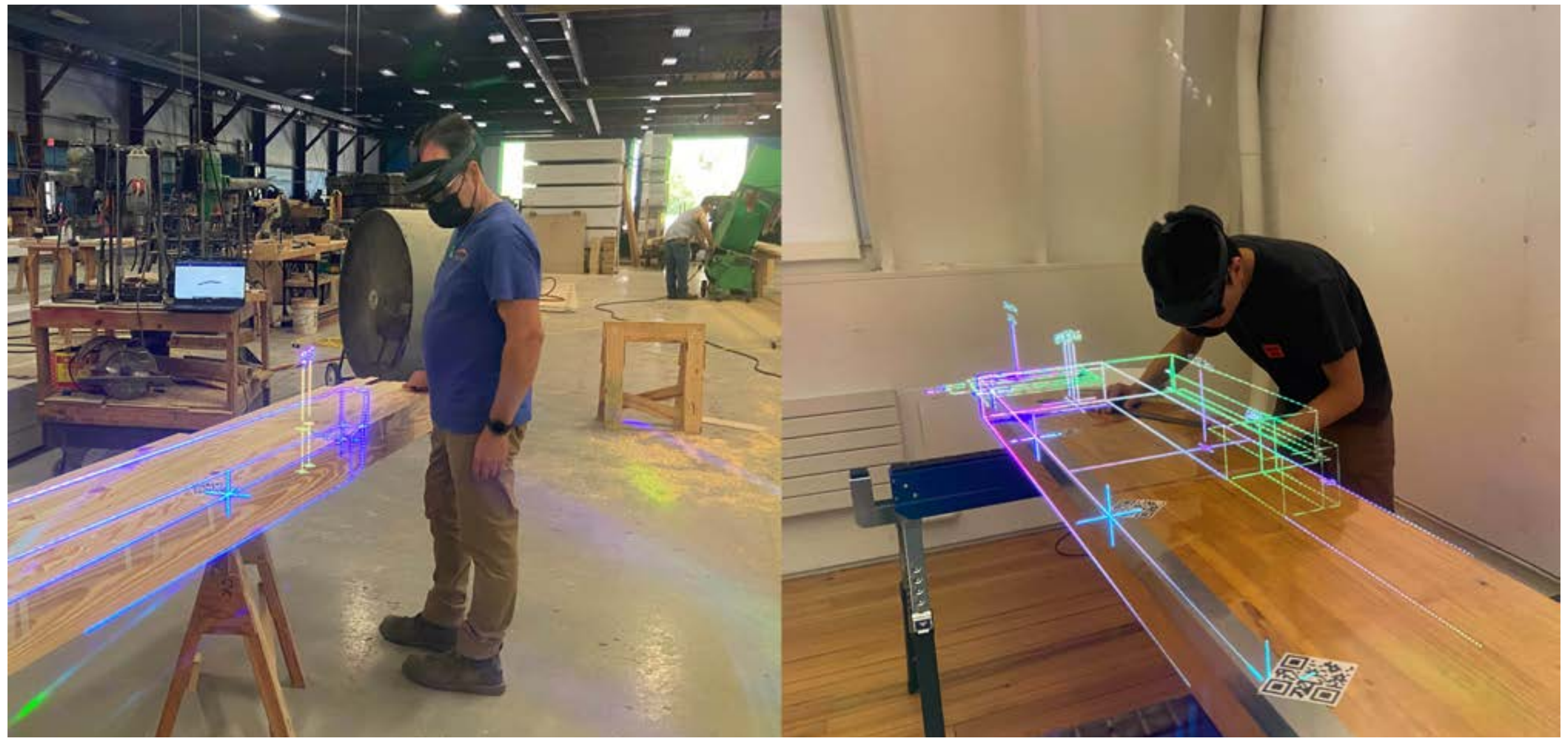

**10** Comparison between factory environment and lab environment

## ACKNOWLEDGMENTS

This work was supported by Unalam (Craig Van Cott, Leif Van Cott, Rik Vandermeulen), Fologram Pty Ltd, and the College of Architecture, Art, and Planning (AAP) at Cornell University. The authors of this paper would like to thank the Robotic Construction Laboratory RCL project team (Lawson Spencer, Chi Zhang, and Lauren Franco) for their valuable feedback and assistance with the experiments.

## REFERENCES

Chi, Hung-Lin, Shih-Chung Kang, and Xiangyu Wang. 2013. "Research Trends and Opportunities of Augmented Reality Applications in Architecture, Engineering, and Construction." Automation in Construction 33 (August):116–22. https://doi.org/10.1016/j.autcon.2012.12.017

Jones, Dennis, and Christian Brischke, eds. 2017. "Wood as Bio-Based Building Material." In Performance of Bio-Based Building Materials, 21–96. Woodhead Publishing. https://doi.org/10.1016/B978-0-08-100982-6.00002-1

Kyaw, Alexander Htet, Jack Otto, Leslie Lok. 2023. "Active Bending in Physics-Based Mixed Reality: The Design and Fabrication of a Reconfigurable Modular Bamboo System." In Dokonal, W, Hirschberg, U and Wurzer, G (Eds.), Digital Design Reconsidered - Proceedings of the 41st Conference on Education and Research in Computer Aided Architectural Design in Europe (eCAADe 2023) - Volume 1, Graz, 20-22 September 2023, Pp. 169–178. https://doi.org/10.52842/conf.ecaade.2023.1.169

Kyaw, Alexander Htet, Lawson Spencer, and Leslie Lok. 2024. "Human–Machine Collaboration Using Gesture Recognition in Mixed Reality and Robotic Fabrication." In Architectural Intelligence Volume 3 (1): 11. https://doi.org/10.1007/s44223-024-00053-4

Kyaw, Alexander Htet, Lawson Spencer, Sasa Zivkovic, and Leslie Lok. 2023. "Gesture Recognition for Feedback Based Mixed Reality and Robotic Fabrication: A Case Study of the UnLog Tower" In Proceedings of the 5th Conference on Computational Design and Robotic Fabrication https://doi.org/10.1007/978-981-99-8405-3_28

Kyaw, Alexander Htet, Arvin Xu, Gwyllim Jahn, Nick Berg, Cameron Newnham, and Sasa Zivkovic. 2023. "Augmented Reality for High Precision Fabrication of Glue Laminated Timber Beams." In Automation in Construction Volume 152 https://doi.org/10.1016/j.autcon.2023.104912

Song, Yang, Richard Koeck, and Shan Luo. 2021. "Review and Analysis of Augmented Reality (AR) Literature for Digital Fabrication in Architecture" 128 (August):103762. https://doi.org/10.1016/j.autcon.2021.103762

## IMAGE CREDITS

All drawings and images by the authors.

**Alexander Htet Kyaw** is a researcher developing new tools and workflows for human-machine-material collaboration through computational design and fabrication. His work integrates multi-disciplinary topics such as augmented reality, digital fabrication, robotics, biomaterials, deployable structures, simulation, computer vision, and machine learning. He is a dual degree candidate at Massachusetts Institute of Technology (MIT), working towards a Master of Science in Architectural Studies in Computation along with a Master of Science in Electrical Engineering and Computer Science. At MIT, Alexander works with the Digital Structures Lab, Media Lab, and Computer Science and Artificial Intelligence Lab.

**Arvin Xu** was a research associate at the Cornell Robotic Construction Laboratory (RCL) at Cornell College of Architecture,Art, and Planning. He graduated with a Bachelor of Architecture from Cornell University department of Architecture.

**Sasa Zivkovic** is an Assistant Professor at Cornell UniversityAAP, the Director of the RCL at Cornell AAP, and a Co-Principalat HANNAH. Zivkovic's research focuses on the development ofsus-tainable robotic construction technologies, material systems,and fabrication processes .

**Gwyllim Jahn** is a co-founder of Fologram and a Lecturerin Architecture at RMIT University in Melbourne. His work focuses ondesigning for mixed reality fabrication, most notably in the design andconstruction of the 2019 Tallinn Architecture Biennale Pavilion. Gwyllim'sresearch has been published in leading compu-tational design confer-ences and journals including IJAC, ACADIA and RobArch and he hasgiven talks, presentations and workshops at international institutionsincluding MIT, Stuttgart ICD, UCL, SciArc, Tongji and Tsinghua University.

**Cameron Newnham** is the co-founder and CTO of Fologram where heleads the technical development of mixed reality software for the designand construction industry. His experience lies in the creation of noveltools for designing and fabricating complex geometric systems, rangingfrom code libraries to mixed reality interfaces and extending to machinedesign and robotic fabrication. Cameron has experience as a computa-tional designer in internationally renowned architectural practices, andacademic experience as an Associate Lecturer – Industry Fellow at RMITUniversity. Cameron has led numerous international design and buildworkshops in Shanghai, New York, Paris, Boston, Sydney, and Melbourne

**Nick van den Berg** is the co-founder and CEO of Fologram, a designresearch practice and technology startup building a plat-form fordesigning and making in mixed reality. Fologram's platform is being usedby world leading architecture practices, product design houses, manu-facturers and design schools internationally. Nick is especially interestedin building solutions that are utilised by a large user base around theworld and has assisted with workshops focusing on utilising augmentedand mixed reality as a design and fabrication tool at DMS, CAADRIA,Cooper Union, McNeel Europe, UDK & TU Berlin.